\documentclass[conference,a4paper]{APSIPA2018}
\usepackage{multirow}
\usepackage{graphicx}
\usepackage{amsmath}
\usepackage[psamsfonts]{amssymb}
\usepackage{amsxtra}
\usepackage{threeparttable}
\usepackage{booktabs}

\begin{document}

\title{A Capsule based Approach for Polyphonic Sound Event Detection}

\author{%
\authorblockN{%
Yaming Liu, Jian Tang, Yan Song, Lirong Dai
}
\authorblockA{%
National Engineering Laboratory for Speech and Language Information Processing,}
\authorblockA{%
University of Science and Technology of China, Hefei, China \\
\{lym66, enjtang\}@mail.ustc.edu.cn, \{songy, lrdai\}@ustc.edu.cn}
}

\maketitle
\thispagestyle{empty}

\begin{abstract}
Polyphonic sound event detection (polyphonic SED) is an interesting but challenging task due to the concurrence of multiple sound events.
Recently, SED methods based on convolutional neural networks (CNN) and recurrent neural networks (RNN) have shown promising performance.
Generally, CNN are designed for local feature extraction while RNN are used to model the temporal dependency among these local features.
Despite their success, it is still insufficient for existing deep learning techniques to separate individual sound event from their mixture, largely due to the overlapping characteristic of features.
Motivated by the success of Capsule Networks (CapsNet), we propose a more suitable capsule based approach for polyphonic SED.
Specifically, several capsule layers are designed to effectively select representative frequency bands for each individual sound event.
The temporal dependency of capsule's outputs is then modeled by a RNN.
And a dynamic threshold method is proposed for making the final decision based on RNN outputs.
Experiments on the TUT-SED Synthetic 2016 dataset show that the proposed approach obtains an F1-score of 68.8\% and an error rate of 0.45, outperforming the previous state-of-the-art method of 66.4\% and 0.48, respectively. 
\end{abstract}

\section{Introduction}
  Sound event detection (SED), also known as acoustic event detection, aims at detecting the onset and offset times of sound events automatically and giving a label to each event.
  With the help of SED technology, computers can understand the environment around via sound and response to it.
  Recently, SED has received increasing interests due to its promising future with wide range of applications in our daily life, including acoustic surveillance \cite{valenzise2007scream}, bio-acoustical monitoring \cite{stowell2015acoustic} and smart facilities in intelligent buildings \cite{sehili2012sound}. 

  According to whether SED allows multiple sound events to occur simultaneously, it can be categorized into monophonic and polyphonic ones,
  For monophonic SED, there exists a certain pattern for each individual sound event in spectrogram.
  For example, the rain event always fills the entire frequency bands, while the thunder event appears at low frequency bands.
  However, for polyphonic SED task, these patterns are very likely to overlap, which make it difficult to effectively separate individual pattern and make a correct detection for each event.

  Traditional approaches for polyphonic SED include hidden Markov model (HMM) \cite{heittola2013context} and non-negative matrix factorization (NMF) \cite{heittola2011sound}.
  In \cite{heittola2011sound}, NMF is used to separate the audio signal into 4 single tracks, where each track represents a combination of the original sources.
  This can be seen as a coarse separation of sound events.
  Recently, feedforward neural networks (FNN) and convolutional neural networks (CNN) have been successfully applied to audio event classification \cite{mcloughlin2015robust,zhang2015robust} as well as polyphonic SED \cite{cakir2015polyphonic, cakir2016filterbank}.
  Recurrent neural networks (RNN) \cite{parascandolo2016recurrent} have achieved quite good performance by integrating information from the earlier time context.
  In \cite{parascandolo2017convolutional}, the CRNN which combines the strength of both CNN and RNN has obtained state-of-the-art polyphonic SED performance.

\begin{figure*}[t]
\begin{center}
\includegraphics[width=7.1in]{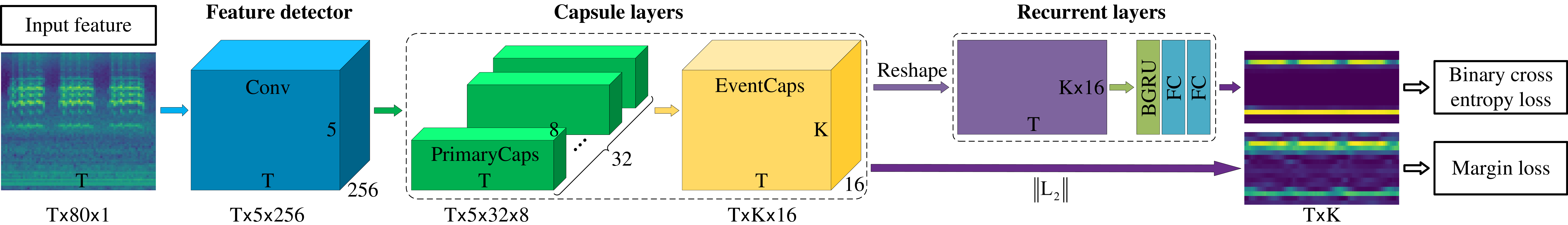}
\end{center}
\caption{Overall architecture of the proposed approach, which consists of three parts. 1) Feature detector: a group of convolutional layers with pooling only on frequency axis. 2) Capsule layers: the outputs of convolutional layers are fed into two capsule layers. 3) Recurrent layers: a bidirectional GRU and two FC layers are used to learn temporal context information and estimate event activity probabilities.}
\label{architecture}
\vspace*{-3pt}
\end{figure*}

  In this paper, we propose a capsule based approach for polyphonic SED, as shown in Fig. \ref{architecture}.
  This is motivated by Capsule Networks (CapsNet) \cite{sabour2017dynamic}, which have shown promising results on highly overlapped digital numbers classification.
  CapsNet are designed to predict the entire entity through partial information and to select suited predictions for the final classification.
  This characteristic may be useful for polyphonic SED task to separate each individual sound event from overlapped features of the mixture.
  In this work, firstly, a stack of convolutional layers are designed to extract local features from the input log mel band energies.
  Then the outputs of CNN are fed into two capsule layers, where local features from different frequency bands and channels are selected to predict multiple objects.
  A RNN is further applied to model the temporal dependency of capsule layers's outputs.
  To learn effective capsule representation, capsule layers and recurrent layers are jointly trained with two different loss functions concurrently.
  Compared with the existing deep learning based polyphonic SED methods, the proposed capsule based approach can effectively select representative frequency bands for each individual sound event, which is more suitable for separating sound events from their mixture.
  The performance is further improved by using a dynamic threshold according to validation metrics.
  Experiments on the TUT-SED Synthetic 2016 dataset show that the proposed approach obtains an improvement over the previous state-of-the-art method.

The main contributions of this study can be summarized as following:
\begin{itemize}
\item A capsule based framework is proposed for polyphonic SED to alleviate the overlap problem.
\item A dynamic threshold strategy is used to make the final classification decision. This is a simple but effective decision method for polyphonic SED task.
\item We experimentally demonstrate the validity of the proposed approach and analyze how events separate from their mixture through visualization.
\end{itemize}


\section{Method}

\subsection{Overview}
The proposed framework illustrated in Fig. \ref{architecture} includes three parts as follows.
1) Feature detector, which is composed of several convolutional layers with pooling only on frequency axis (time axis does not shrink), time-frequency representations of audio signal are fed into the detector.
2) Capsule layers, including a PrimaryCaps and an EventCaps, which are designed to select features from different frequency bands and channels.
3) Recurrent layers, which are used to learn temporal context information and estimate event activity probabilities.
Hyperparameters used are presented in Table \ref{hyperparameters}.

\begin{table*}[t]
  \centering
  \begin{threeparttable}
  \caption{Hyperparameters used in the proposed approach}
  \label{hyperparameters}
    \begin{tabular}{p{30mm}ccccccccc}
    \toprule
    \multirow{2}{*}{}&
    \multicolumn{4}{c}{Feature detector}&\multicolumn{2}{c}{Capsule layers}&\multicolumn{3}{c}{Recurrent layers}\\
    \cmidrule(lr){2-5} \cmidrule(lr){6-7} \cmidrule(lr){8-10}
    &Conv1&Conv2&Conv3&Conv4&PrimaryCaps&EventCaps&GRU&FC&FC \\
    \midrule
    kernel&256@$3\times3$&256@$3\times3$&256@$3\times3$&256@$3\times3$&-&-&-&-&- \\
    stride&$1\times1$&$1\times1$&$1\times1$&$1\times1$&$1\times1$&-&-&-&- \\
    pooling size&-&$1\times4$&$1\times2$&$1\times2$&-&-&-&-&- \\
    activation function&ReLU&ReLU&ReLU&ReLU&squashing&squashing&-&ReLU&Sigmoid \\
    num of hidden units&-&-&-&-&-&-&256&512&16 \\
    dim of capsule&-&-&-&-&8&16&-&-&- \\
    \bottomrule
    \end{tabular}
  \end{threeparttable}
\end{table*}

\subsection{Feature detector}
We use 4 convolutional layers to detect local features in this work.
Max-pooling is used to reduce frequency dimensionality, while time axis keeps the same to match the length of target.
Log mel band energies $ \mathcal X \in \mathbb{R} ^{F \times T} $ is fed into the feature detector with zero-padding, where $F$ is frequency bins of input features, $T$ is the number of frames in a sample.
The output of feature detector is a tensor $ \mathcal H \in \mathbb{R} ^{M \times F^{'} \times T} $, where $M$ is the number of feature maps in the last convolutional layer, $ F^{'} $ is the number of frequency bands after series of pooling operations.

\subsection{Capsule layers}
Capsules are vectors whose dimensions are associated with various properties of objects, such as location, size, orientation, etc.
The length of each vector represents the activity probability of a specific object, and is limited to range from 0 to 1 by a nonlinear squashing function in \eqref{eq_3}.
Two capsule layers are used in this work, a PrimaryCaps and an EventCaps.
PrimaryCaps is a convolutional capsule layer with 32 channels.
Each channel consists of 8D capsules.
These capsules are also called low-level capsules which are fed into EventCaps later to obtain high-level ones.
In EventCaps, firstly, prediction vectors of high-level capsules are calculated by multiplying outputs of low-level capsules by a weight matrix, as in \eqref{eq_1}.
Then these prediction vectors are selected by routing-by-agreement process according to similarity between each high-level capsule's output and its prediction vectors, as in \eqref{eq_2}\eqref{eq_4}\eqref{eq_5}.
The more similar a prediction vector and its corresponding high-level capsule's output are, the larger the connection weight between them is.
This weight gain further increases the contribution of that prediction vector to its corresponding high-level capsule.

Let $\boldsymbol u_{i}$ denote the output of low-level capsule $i$, and $\boldsymbol v_j$ is the output of high-level capsule $j$, then $\boldsymbol v_j$ can be calculated as following
\begin{align}
& \hat {\boldsymbol u}_{j|i} = \boldsymbol W_{ij} \boldsymbol u_{i} \label{eq_1} \\
& \boldsymbol s_j = \sum_{i} c_{ij} \hat {\boldsymbol u}_{j|i} \label{eq_2} \\
& \boldsymbol v_j = \frac{\lVert \boldsymbol s_j \rVert ^2}{1+\lVert \boldsymbol s_j \rVert ^2} \frac {\boldsymbol s_j}{\lVert \boldsymbol s_j \rVert} \label{eq_3}
\end{align}
where $ \hat {\boldsymbol u}_{j|i} $ is the prediction vector of high-level capsule $j$ from low-level capsule $i$, $ \boldsymbol W_{ij} $ is the corresponding weight matrix.
The coupling coefficients $ c_{ij} $ are  determined by dynamic routing process as follows
\begin{align}
& c_{ij} = \frac{{\rm exp}(b_{ij})}{\sum_k {\rm exp}(b_{ik})} \label{eq_4} \\
& b_{ij} \leftarrow b_{ij} + \hat {\boldsymbol u}_{j|i} \cdot \boldsymbol v_j \label{eq_5}
\end{align}
where $b_{ij}$ are log prior probabilities that low-level capsule $i$ should be coupled with high-level capsule $j$.
The $b_{ij}$ are initialized to 0 and updated by the similarity between prediction vector $\hat {\boldsymbol u}_{j|i}$ and high-level capsule's output $\boldsymbol v_j$.
This similarity can be measured by a scalar product operation.

Finally, margin loss of each output capsule $k$ is calculated
\begin{multline}
L_k = T_k {\rm max} (0,m^{+} - \lVert v_k \rVert)^2 + \\
\lambda (1-T_k) {\rm max} (0, \lVert v_k \rVert - m^{-})^2
\end{multline}
where $T_k$ is 1 when class $k$ actually exists, otherwise 0.
Terms $m^+$, $m^-$ and $\lambda$ are hyperparameters, which are set to the same values as origin CapsNet \cite{sabour2017dynamic}.
The total loss is the sum of the losses of all output capsules.

In this work EventCaps calculates results on each frame simultaneously.
The main process of capsule layers part is described as follows.

\begin{itemize}
\item PrimaryCaps, each channel contains $ F^{'} \times T $ 8D capsules, i.e., $ F^{'} \times 32 $ primary capsules for each frame.
\item Capsules of each frame are fed into EventCaps to compute $K$ 16D capsules, where $K$ is the number of event classes. Each 16D capsule represents one sound event. The output of EventCaps is a tensor $ \mathcal J \in \mathbb{R} ^{16 \times K \times T} $.
\item Calculating length of each capsule inside every frame. The output is a tensor $ \mathcal K \in \mathbb{R} ^{K \times T} $.
\end{itemize}

\subsection{Recurrent layers}
A RNN is used to learn temporal context information, since the temporal dependency has proved to be important in sound event analysis task \cite{parascandolo2016recurrent,parascandolo2017convolutional,xu2017convolutional}.
We reshape the output tensor of DigitCaps $ \mathcal J \in \mathbb{R} ^{16 \times K \times T} $ to $ \mathcal M \in \mathbb{R} ^{(16 \times K) \times T} $, i.e. combining $K$ 16D capsules in each frame.
These combined vectors are fed into a bidirectional gated recurrent unit (GRU). The bidirectional GRU outputs hidden state $\mathbf h_t$ at each frame $t$, followed by a feedforward layer with ReLU activation function.
Finally, another feedforward layer with sigmoid activation function is used as the output layer.
The output of recurrent layers is a tensor $ \mathcal F \in \mathbb{R} ^{K \times T} $, representing events activity probabilities of $K$ sound events along $T$ frames.

\subsection{Dynamic Threshold}
Event activity probabilities are binarized by a threshold to obtain binary representation.
The threshold can be very crucial to the performance of polyphonic SED system since we don't know how many events exist in one frame.
In our work, we select an optimal threshold $C_{opt}$ between [0.5, 0.9] for each model based on the validation metrics, the optimal threshold is used on the test set afterwards.

\section{Experiments}

\subsection{Datasets and Metrics}
We evaluate the proposed approach on the dataset TUT-SED Synthetic 2016 \cite{parascandolo2017convolutional}. 
Segment-based error rate (ER) and F1-score (F1) proposed in \cite{mesaros2016metrics} are used as evaluation metrics in this work.
Intermediate statistics are accumulated over the segments of the whole test set and then used to calculate ER and F1, which is called micro-averaging.
To ensure comparability with the baseline system \cite{parascandolo2017convolutional}, we use two kinds of segments length, single frame (40ms) and one-second. Thus four evaluation metrics are used in this work, i.e., $ER_{frame}$, $F1_{frame}$, $ER_{second}$ and $F1_{second}$, where $ER_{frame}$ and $F1_{frame}$ as primary evaluation metrics.

\subsection{Baseline}

We compare our work with the previous state-of-the-art approach CRNN \cite{parascandolo2017convolutional}.
Work in \cite{parascandolo2017convolutional} uses a CNN as feature extractor, log mel band energies is fed into the CNN.
Time axis keeps the same during convolution, and feature maps are stacked along the frequency axis afterwards.
These stacked feature maps are fed into a GRU later.
Feedforward layer with sigmoid activation function is used as the output layer to obtain sound event activity probabilities of each frame.

\subsection{Experiment Setting}
%
We use the same log mel band energies as \cite{parascandolo2017convolutional}, except that the 80 mel bands are used instead.
Each mel band is normalized by subtracting its mean and dividing by its standard deviation calculated over the training set.
The normalized log mel band energies are split into samples by a sliding window with fixed length $T$ (frames, 128 in this work).
Samples are overlapped during training and are nonoverlapping during validation and test, which is also the same as \cite{parascandolo2017convolutional}.

All the networks are trained with Adam \cite{kinga2015method} optimizer with a fixed learning rate of 0.0001.
Dynamic threshold strategy is used in all our networks unless specifically mentioned. Batch normalization and dropout with dropout rate 0.25 are used after each convolutional layer.
We use early stopping during the training process, holt the training if validation metrics are not improving for more than 10 epochs.
The model with the best performance on validation set is chosen as the final model which is used to calculate results on test set.
Each experiment is repeated 10 times with different random seeds. 

The six comparison systems are as follows:

\textbf {CNN} is the first baseline from work \cite{parascandolo2017convolutional}, it consists three convolutional layers and no recurrent layer.

\textbf {CRNN} is the second baseline from work \cite{parascandolo2017convolutional}, including three convolutional layers and one GRU recurrent layer.

\textbf {CapsNet} is the CapsNet baseline which only includes feature detector and capsule layers. The model regards lengths of capsules of DigitCaps $\mathcal K$ as outputs. Margin loss is calculated between $\mathcal K$ and ground truth $ \mathcal Y \in \mathbb{R} ^{K \times T} $.

\textbf {Capsule-RNN} is our proposed approach as described in Section \uppercase\expandafter{\romannumeral2}. During the training process, two losses are calculated, i.e., binary cross-entropy loss between outputs of recurrent layers $\mathcal F$ and ground truth $ \mathcal Y$, margin loss between capsules's lengths $\mathcal K$ and ground truth $ \mathcal Y $. The final loss is a weighted sum of these two losses by a couple of weights [0.7, 0.3]. During the validation and test process, outputs of recurrent layers $\mathcal F$ are considered as event activity probabilities, and are binarized by dynamic threshold (DT).

\textbf {Capsule-RNN without DT} use a fixed threshold $C=0.5$ (same as \cite{parascandolo2017convolutional}) in Capsule-RNN to binarize event activity probabilities.

\subsection{Results}

In this section, we provide the mean and the standard deviation of F1 and ER in all experiments described above.
As presented in Table \ref{main results}, CapsNet improves frame-based F1 and ER by relative 8\% and 11\% respectively compared with CNN baseline.
Meanwhile, Capsule-RNN achieves a relative improvement of 4\% on frame-based F1 and 6\% on frame-based ER over CRNN baseline.
Considering the number of parameters used for Capsule-RNN and CRNN are similar (CapsNet is less than CNN), these improvements indicate an architectural advantage of capsule based methods compared with CNN based methods.
On the other hand, compared with CapsNet, Capsule-RNN obtains a further 7\% and 10\% relative improvement on frame-based F1 and ER respectively, implying that the temporal context information is important to polyphonic SED task.
This is consistent with previous work.
Comparison of Capsule-RNN and Capsule-RNN without DT shows that a proper decision threshold is essential for ER.


\begin{table}
\centering
\begin{threeparttable}
\caption{ER and F1 of one frame segment based and one second segment based for baselines and proposed models}
\label{main results}
\begin{tabular}{lcccc}
\toprule
Model & $F1_{frame}$ & $ER_{frame}$ & $F1_{second}$ & $ER_{second}$ \\
\midrule
CNN \cite{parascandolo2017convolutional} & $\text{59.8}\pm\text{0.9}$ & $\text{0.56}\pm\text{0.01}$ & $\text{59.9}\pm\text{1.2}$ & $\text{0.78}\pm\text{0.08}$ \\
CapsNet & $\text{64.6}\pm\text{0.9}$ & $\text{0.50}\pm\text{0.01}$ & $\text{65.0}\pm\text{0.6}$ & $\text{0.62}\pm\text{0.01}$ \\
\midrule
CRNN \cite{parascandolo2017convolutional} & $\text{66.4}\pm\text{0.6}$ & $\text{0.48}\pm\text{0.01}$ & $\text{68.7}\pm\text{0.7}$ & $\text{0.47}\pm\text{0.01}$ \\
Capsule-RNN & $\textbf{68.8}\pm\textbf{0.7}$ & $\textbf{0.45}\pm\textbf{0.01}$ & $\textbf{69.2}\pm\textbf{0.5}$ & $\textbf{0.45}\pm\textbf{0.01}$ \\
Capsule-RNN \tnote{a} & $\text{68.6}\pm\text{0.8}$ & $\text{0.47}\pm\text{0.02}$ & $\text{68.1}\pm\text{0.8}$ & $\text{0.51}\pm\text{0.03}$ \\
\bottomrule
\end{tabular}
\begin{tablenotes}
\item[a] Capsule-RNN without DT
\end{tablenotes}
\end{threeparttable}
\end{table}

\begin{figure}[t]
\centering
\includegraphics[width=1.0\linewidth]{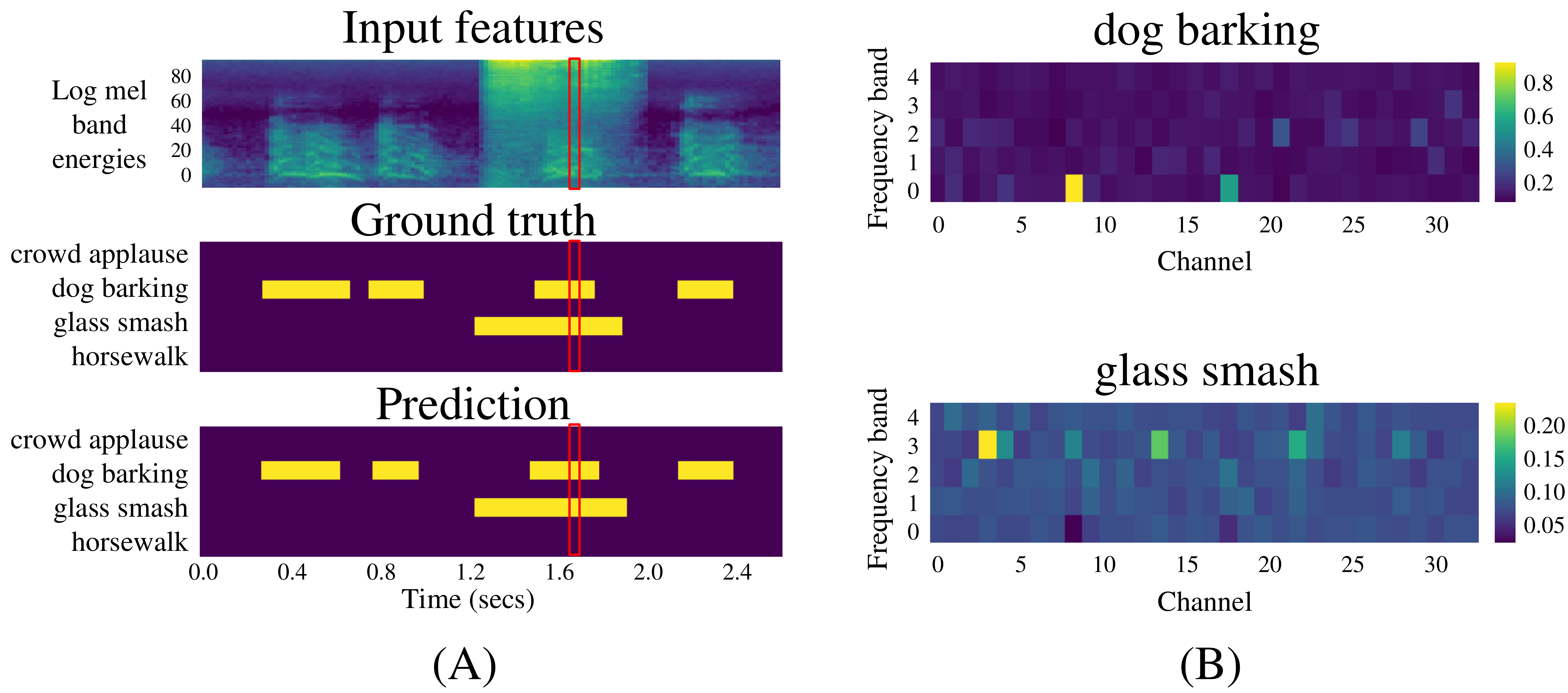}
\caption{(A) Input features, ground truth and prediction of an example from test set. (B) Coupling coefficients $c_{ij}$ at the 80th frame. Each point in the image refers to the coupling coefficient between a high-level capsule (\textit{dog barking} or \textit{glass smash}) and its prediction vector from a low-level capsule. The vertical axis can be considered as frequency bands between 0-22050 Hz (0 for lowest frequency band, 4 for highest frequency band), while the horizontal axis represents the channels.}
\label{Sample}
\vspace*{-3pt}
\end{figure}

Fig. \ref{Sample} shows an example from the test set for predicting events activities along 128 frames (2.56s), which also illustrates how the proposed approach is able to distinguish events from their mixture.
We draw the coupling coefficients $c_{ij}$ at the 80th frame (circled in red box) in Fig. \ref{Sample}-B. 
This distribution of coupling coefficients can be seen as a selection of frequency bands and channels when producing a high-level capsule from its prediction vectors. 
Two sound events are contained at the 80th frame, where \textit{dog\_barking} presents at middle and low frequency bands, and \textit{glass\_smash} covers all bands.
Fig. \ref{Sample} demonstrates that the proposed approach successfully detects these two events from their mixture by selecting channels on different frequency bands for \textit{dog\_barking} (the lowest frequency band) and for \textit{glass\_smash} (the second-highest frequency band), respectively.

\section{Conclusions}

In this paper, we proposed a capsule based approach for polyphonic sound event detection to alleviate the overlap problem.
In our approach, several capsule layers were designed to effectively select representative frequency bands for each individual sound event.
Also, the dynamic threshold strategy was proposed for selecting an optimal threshold for each model.
Experiments showed that the proposed approach outperformed the previous state-of-the-art CRNN method.

\section*{Acknowledgment}
This work was supported by National Key R\&D Program of China (Grant No. 2017YFB1002202) and the Key Science and Technology Project of Anhui Province (Grant No. 17030901005).

\bibliographystyle{IEEEtran}

\bibliography{mybib}
\end{document}